\documentclass[prb,twocolumn]{revtex4}
\usepackage{color,soul}
\usepackage {graphicx}
\usepackage {epstopdf}

\begin{document}
\title{Role of f-d exchange interaction and Kondo scattering in Nd doped pyrochlore Iridate
 (Eu$_{1-x}$Nd$_x$)$_{2}$Ir$_{2}$O$_{7}$ } 

\author{Sampad Mondal$^{a,b,c}$\footnote{Email:sampad100@gmail.com}, M. Modak$^b$, B. Maji$^{b,d}$, M. K. Ray$^e$, S. Mandal$^b$, Swapan K. Mandal$^a$, M. Sardar$^f$ and S. Banerjee$^b$\footnote{Email:sangam.banerjee@saha.ac.in}}
\address{$^a$ Department of Physics, Visva-Bharati, Santiniketan 731235, India\\
	$^b$ Saha Institute of Nuclear Physics, HBNI, 1/AF Bidhannagar, Kolkata 700064, India\\
	$^c$Ramsaday College, Amta, Howrah 711401, India\\
	$^d$ Acharya Jagadish Chandra Bose College, 1/1B, A. J. C. Bose Road, Kolkata 700020, India\\
	$^e$ Institute of Solid State Physics, University of Tokyo, Kashiwa 277-8581, Japan\\
	$^f$Material Science Division, Indira Gandhi Centre for Atomic Research, Kalpakkam, India}
\begin{abstract}
	We report study of magnetization, resistivity, magnetoresistance and specific heat of the pyrochlore
	Iridate  (Eu$_{1-x}$Nd$_x$)$_{2}$Ir$_{2}$O$_{7}$ with x=0.0, 0.5 and 1.0,
	where spin orbit coupling, electronic correlation, 
	magnetic frustration and Kondo scattering coexists.
	Metal insulator transition temperature ($T_{MI}$) decrease with increase in Nd content but always coincides with
	magnetic irreversibility temperature (field induced moment).
	Resistivity below $T_{MI}$ do not fit with either activated (gap) or to any power law (gapless) dependence. The Curie constant
	show surprising result, that Nd induces singlet correlation (reduction of para-moment) in Ir sublattice.
	Magnetoresistance is negative at low temperatures below 10 K and 
	increases strongly with increase in x and vary quadratically with field switching over to linear dependence
	above 50 kOe. 
	Low temperature specific heat shows Schottky peak, coming from Nd moments, showing 
	existence of doublet split in Nd energy level, arising from f-d exchange interaction. All materials show presence of a linear specific heat
	in the insulating region. The coefficient of linear specific heat for x = 0.0 does not vary with external
	magnetic field but varies superlinearly for x = 1.0 materials.
	We argue that linear specific heat  probably rules out 
	weakly correlated phases like Weyl fermions.
	We propose that with the introduction of Nd at Eu site the system evolves from chiral spin liquid  
	with gapless spinon excitations with a very small charge gap to Kondo type interaction superposed on chiral spin liquid coexisting with long range antiferromagnetic ordering.  Huge increase of 
	magnetoresistance with increase in Nd concentrations shows importance of Kondo scattering in the chiral spin liquid material by rare earth 
	moments.

	\end{abstract}
\maketitle
{\large\bf{Introduction}}
\vskip 0.2cm
Focus in condensed matter physics now a days is in understanding emergence of Dirac-like fermions under various 
circumstances.\cite{vish}
A common theme is spin-orbit entanglement produced by spin-orbit interaction,  influencing electronic 
and magnetic transitions.\cite{krempa}
Topology of the band structure is also important and it was suggested that in the weak correlation regime, 
quadratic band touching
at some points in the Brilloiun zone can lead to a large number of phases related to topological insulators, 
like three-dimensional Dirac\cite{dirac}
and Weyl\cite{we} semimetals and  axion insulator.\cite{axon} Specifically in the magnetically ordered phase, where
time reversible symmetry is broken but inversion symmetry is preserved, the 
pairs of quadratic band touching points lead to linearly dispersing
Dirac fermion type spectrum, along with definite chirality, i.e. $\bar k \cdot \bar \sigma = \pm 1$, 
where $\bar k$ and $\bar \sigma$ are unit vectors along
the momentum and spin of electrons. This leads to many interesting properties like Hall effect without external 
magnetic field, 
negative/positive magnetoresistance when
electric field and magnetic fields are parallel/perpendicular.\cite{Takumi,Son,Machida}

In the strong correlation limit, the already narrow bands due to spin-orbit coupling, may open up gap. With increase in 
local coulomb correlation, 
pairs of opposite chirality Weyl points move
towards Brillouin zone boundary and annihilate pairwise to open up a gap, forming  spin-orbit assisted Mott insulators.\cite{mott}
Owing to frustration in the magnetic exchange interaction (like in pyrochlores), one might get either metallic or 
insulating spin-liquid phases.\cite{spinliq}  This spin liquid phase, may also be chiral, but this time chirality is defined in 
terms of real space localised spin variables
like $S_i \cdot S_j \times S_k$ developing non zero expectation values,  where $S_i,S_j, S_k$ are spins at 
sites $i,j,k$ in a triangular plaquette.
Interestingly this phase can also give anomalous Hall effect (without external magnetic field) and negative magnetoresistance 
like Weyl fermions if the transport gap is small.

To  study the combined  effect of  spin-orbit coupling and electronic correlation,
the $5d$ iridium oxides, like pyrochlore iridates Ln$_2$Ir$_2$O$_7$ where Ln is a lanthanide
are the best candidates. 
The interpenetrating corner sharing tetrahedral structure is favourable to form a narrow flat band that enhances the 
effect of electron correlation and spin orbit coupling (SOC), and both of these energies  are comparable to the band width. 
They also have a structure where antiferromagnetic interaction between magnetic ions is frustrated. These materials 
are proposed\cite{pro}
to have many interesting topological phases and is being persued by experimentalists to discover 
interesting properties of matter.

Ln$_2$Ir$_2$O$_7$ (R=rare earth) have shown many interesting  transport and magnetic 
properties, \cite{Zhang,Daiki,MATSUHIRA1} and shows a transition from incoherent and strongly correlated 
metal (at high temperature)
to an antiferromagnetically ordered insulator/semimetal (at low temperature) except when Ln=Pr. Transition temperature
decreases as the  ionic 
radius of the rare earth atom\ increases.\cite{MATSUHIRA1} Metal to insulator/semimetal transition is reminescent 
of Mott-Hubbard transition
but with the added complexity, that (1) spin and orbital degrees of freedoms are entangled due to spin-orbit 
coupling, (2) both
transition metal and rare earth sublattices are magnetically frustrated, and (3) there is local 
Kondo coupling (f-d exchange interaction)
between localised rare earth
moments and the transition metal electrons.
In  Eu$_{2}$Ir$_{2}$O$_{7}$  we can avoid 
f-d exchange interaction because net moment of Eu$^{3+}$ is zero. This is a  suitable system to study the effect of both SOC 
and electronic correlation. 
A continuous phase transition from paramagnetic metal to antiferromagnetic insulating state was 
observed in single crystals.\cite{Ishikawa} Low frequency optical conductivity\cite{sarma} increases linearly 
with frequency. This is
consistent with a Dirac-like spectrum with density of states varying linearly with energy, suggesting 
that Eu$_2$Ir$_2$O$_7$ is a Weyl Semimetal. From analysis of resistivity, Tafti et al.\cite{Tafti} 
have suggested that Eu$_2$Ir$_2$O$_7$ goes from paramagnetic metal to Weyl semimetal (in an intermediate temperture window)
and finally to an antiferromagnetic insulator at the lowest temperture.  Paradoxically in the low temperature region (below 10 K) the optical 
conductivity was also found to be proportional to frequency.
When the rare earth sites have moment (like Nd) the f-d exchange interaction (Kondo) 
can mediate RKKY type exchange interaction between the localized  rare earth moments, which along with 
superexchange interaction
between themselves, may  induce the 
ordering of Nd sub lattice system. The magnetic ordering of Nd sub system can in turn modify the 
electronic structure of the Ir sub system, 
promoting 
many exotic quantum correlated phenomena. Even when Nd moments are not ordered yet, the Kondo scattering 
between f and d electrons
can profoundly affect the electronic properties of Ir electrons. Weyl fermion  is a weak correalation concept and are 
unlikely to survive with increase in correlation. It has been argued\cite{Chen}  that f-d interaction 
along with strong correlation can again stabilize  Weyl Semimetal phase.
Polycrystalline 
Nd$_2$Ir$_2$O$_7$ shows  metal insulator transtion at 33 K.\cite{MATHSUHIRA2} Raman scattering  confirms that there is no 
structural change accompanying metal insulator transition.\cite{Hasegawa}
Optical,\cite{ueda}  transport\cite{tian}
and photoemission\cite{nakayama}
experiments suggest a gapped insulating ground state in Nd$_2$Ir$_2$O$_7$. 
Nd$_2$Ir$_2$O$_7$ compound develops a long range 
antiferromagnetic all-in-all-out (AIAO) ordering of the Nd moments\cite{Guo} below 15 K (though the magnitude of ordered Nd moments is
very small), confirmed by 
inelastic neutron  scattering measurement.\cite{Tomiyasu} Long range antiferromagnetic ordering is 
destroyed by applied magnetic field, producing  a field induced insulator to metal transition. \cite{Udea,tian}
We have prepared polycrystalline  Eu$_{2(1-x)}$Nd$_{2x}$Ir$_2$O$_7$ with x=0.0, 0.5 and 1.0, and have studied resistivity, 
magnetoresistance, magnetization and specific heat with and without magnetic field. 

\vskip 0.5cm
{\large\bf{Experimental details}} 
\vskip 0.2cm
All the polycrystalline samples were prepared by solid state reaction method. High purity ingredient powder Eu$_{2}$O$_{3}$, IrO$_{2}$ and Nd$_{2}$O$_{3}$ were mixed in stochiometric ratio and ground well. After pressing the mixture powder in pellet form, heated at 1373 K for 3 days with several intermediate grinding. All the samples were characterised by powder X-ray diffraction (XRD). The room temperature XRD measurement was taken by X-ray diffractometer with Cu K$\alpha$ radiation. Structural parameter were determined using standard Rietveld technique with Fullprof software package. Magnetic measurement was taken by Superconducting Quantum Interference Device Magnetometer (SQUID-VSM) of Quantum Design in the temperature range 3 K - 300 K. Electrical, magnetic and thermal transport measurements were carried out by Physical Properties Measurement System (PPMS). 
\begin{figure} 
	\centering
	\includegraphics[width= 8 cm]{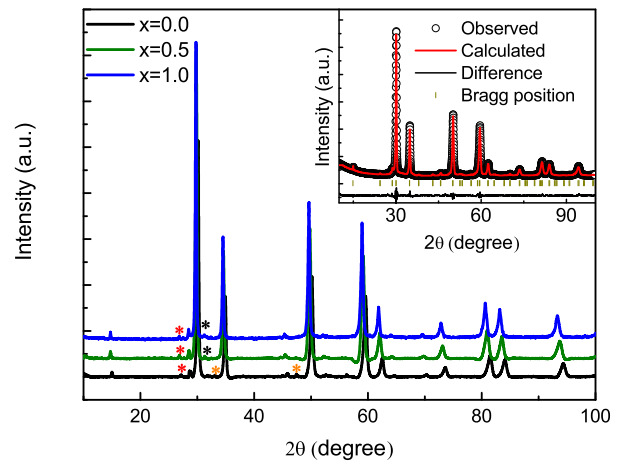}
	\caption{Room temperature X-ray diffraction pattern for the sample  (Eu$_{1-x}$Nd$_x$)$_{2}$Ir$_{2}$O$_{7}$ where x=0, 0.5, 1.0 and  red, orange and black stars are indicating the impurity phase due to non-reacting IrO$_2$, Eu$_2$O$_3$ and Nd$_2$O$_3$
	respectively. Inset fig. shows Rietveld refinement for x=0.0 compound, where scattered data are observed and solid line is fit to the data.}
	\label{XRD}
\end{figure}
\vskip 0.5cm
{\large\bf{Experimental results }}
\vskip 0.2cm
Room temperature XRD pattern for the samples (Eu$_{1-x}$Nd$_x$)$_{2}$Ir$_{2}$O$_{7}$ with 
x=0.0, 0.5, 1.0 are 
shown in fig. \ref{XRD}. There is no modification of the XRD pattern but peaks shift 
to lower angles with Nd 
substitution at Eu site. Observed data of all  samples are refined on the basis of cubic 
structure with 
space group Fd-3m by Rietveld refinement method. Refinement of the x=0.0 compound
is shown in the
inset of fig. \ref{XRD}. All the samples are nearly in pure phase except some minor impurity phases (non-reacting oxide) which are marked by star. In x=0.0 compound, impurity phases due to non-reacting oxide IrO$_2$ and Eu$_2$O$_3$ are 0.62$\%$ and 0.89$\%$ respectively and  IrO$_2$, Nd$_2$O$_3$ impurity phases in  x=0.5 and 1.0 compounds are 0.72$\%$, 0.69$\%$ and 0.78$\%$, 0.82$\%$  respectively.
Obtained lattice 
parameters from Rietveld refinement for x=0, 0.5 and 1.0 compounds are 10.3142 $\AA$, 
10.3472 $\AA$ and 
10.3858 $\AA$ respectively. The increase of lattice parameter with Nd doping indicates 
that Eu$^{3+}$ 
is substituted by larger ionic radius Nd$^{3+}$. 

  \begin{figure} 
    \centering
      	\includegraphics[width= 8 cm]{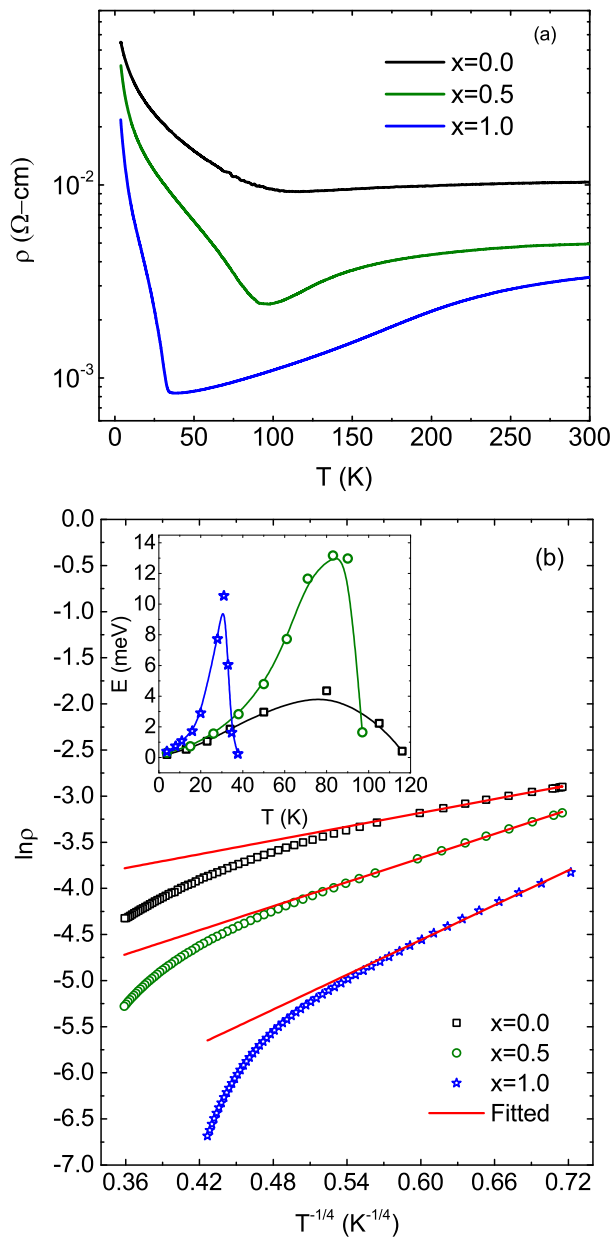}
      	\caption{(a) Temperature dependent resistivity measured with temperature range 3.8-300 K and (b)  Fitting of resistivity below MIT for the compounds (Eu$_{1-x}$Nd$_x$)$_{2}$Ir$_{2}$O$_{7}$ where x=0, 0.5 and 1.0 and inset shows variation of charge gap (E) with temperature for all the samples.}
      	\label{Rho-T_VRHfit}
      \end{figure}
      \vskip 0.5cm
      Fig. \ref{Rho-T_VRHfit}(a) shows temperature dependent resistivity of the sample 
      (Eu$_{1-x}$Nd$_x$)$_{2}$Ir$_{2}$O$_{7}$ 
      with  x=0, 0.5 and 1.0 from 3 K to 300 K. All samples show metal to insulator transition (MIT) i.e. 
      $\frac{d\rho}{dT}$  
      changes from positive to negative  below the MIT temperature ($T_{MI}$). Observed MIT temperatures of 
      x=0, 0.5 and 1.0 compounds 
      are 120 K, 95 K and 35 K respectively. Both $T_{MI}$ 
      as well as  the  resistivity in the
      entire temperature range decreases with Nd doping. This is consistent with earlier 
      observations \cite{MATSUHIRA1} that $T_{MI}$
      decreases with increase in rare earth ionic size. From Rietveld refinement we find  higher ionic radii 
      ion Nd doped at Eu site increases the bond angle between Ir-O-Ir which increases the Ir-O orbital overlap 
      leading to increase Ir-Ir 
      hopping matrix element, i.e.  Ir  t$_{2g}$ band width.\cite{koo} The MI transition gives the 
      appearance of a straightforward Mott-Hubbard
      transition (competition between hopping matrix element and Local Hubbard repulsion), but attempt to 
      fit the resistivity in the
      insulating phase with activated form (with a gap) fails, as it shows that the gap is temperature dependent with a maximum below $T_{MI}$  and very small near 3 K and $T_{MI}$ (see inset fig. \ref{Rho-T_VRHfit}(b)). On the other hand $\mu$SR experiments 
      in Eu and Nd
      compounds show continuous rise of well defined muon precession frequency below $T_{MI}$,\cite{Eu,Nd} indicating 
      long range magnetic ordering into
      a commensurate structure, as is expected in an antiferromagnetically ordered  Mott insulating state. 
      Antiferromagnetic  ordering can emerge in
      weak coupling (small Hubbard U) theories like in Slater antiferromagnetic state also but the 
      temperature dependence of transport gap is disturbing. It has been suggested theoretically\cite{weyl} 
      that arbitrarily small antiferromagnetic ordering of Ir electrons can convert quadratic band touching 
      points into a Weyl semimetal phase with Dirac cone
      linear dispersions at some points on the Brillouin zone. Optical conductivity in Eu$_2$Ir$_2$O$_7$ 
      measured at 7 K shows\cite{sarma}
      a liner dependence on frequence at the lowest frequencies, consistent with  density of states 
      $\rho(E) \propto |E-E_F| $ expected 
      out of a Weyl semimetal. This should also mean that the low temperature resistivity must vary with 
      temperature as ${1 \over T}$ and was predicted
      theoretically.\cite{vish} We find that resistivity of none of the samples varies in a power law (${1 \over T^{\alpha}}$ ) 
      fashion from 3 K to  $T_{MI}$. Single crystal resistivity measurement\cite{Ishikawa} of x=0.0 
      showed a surprising result i.e., the temperature dependence of resistivity of all single crystals (made in 
      the same batch) coincides  above $T_{MI}$, while  below $T_{MI}$  there is wide discrepancies in resistivity values between 
      samples, and the residual resistivity at the lowest temperature varies by four orders of 
      magnitude between samples. There are two sources of possible disorder, (1) slight difference in oxygen
      concentration\cite{Ishikawa} and (2) interchange of a small fraction of Eu and Ir sites.\cite{Telang}
      Both has the net effect of doping hole in the half filled Ir sub system. It is found that more stoichiometric 
      materials have
      larger residual resistivity.\cite{Ishikawa} Sr doping in  Eu$_2$Ir$_2$O$_7$\cite{banerjee} and Ca doping in  
      Nd$_2$Ir$_2$O$_7$\cite{zach} 
      leads to reduction of $T_{MI}$ and both systems above $T_{MI}$ shows non fermi liquid properties. More importantly 
      it is found\cite{eric}
      that Nd$_2$Ir$_2$O$_7$ fragments into magnetic domains separated by domain walls, and the domain wall region 
      has much lower
      resistivity than within the magnetic domains, i.e. the measured conductivity is that of the domain walls and not of bulk. 
      Assuming the antiferromagnetic
      regions as topological insulators, the domain walls might have  gapless surface states. It is to be noted that topological insulator
      and Weyl metal/semimetals arise due to chirality in momentum space.
      Moreover optical\cite{ueda} and transport\cite{tian} 
      and photoemission\cite{nakayama}
      experiments suggest a gapped insulating ground state in Nd$_2$Ir$_2$O$_7$. 
      In view of the fact that there is unavoidable non-stoichiometry as well as domain formation in these materials, 
      it is worthwhile to look at
      theoretical\cite{hai} analysis of disordered Weyl metals. Main conclusions from theoretical calculations are,
      (1) resistivity crosses over to ${1\over T^{0.5}}$ from ${1\over T}$ (for clean Weyl metal), and (2) the 
      magnetoresistance at
      low field is positive (anti-localisation, because of spin orbit coupling) and switches over to negative  at higher field. 
      Both of these are not observed in our materials. 
      
      \begin{table}
      	
      	\centering \caption{Fitting parameter T$_{0}$ and temperature range obtained from VRH model for the sample (Eu$_{1-x}$Nd$_x$)$_{2}$Ir$_{2}$O$_{7}$ }

      	\begin{tabular}{ccc}
      		\hline Sample&\hspace{0.5cm}Temperature range (K)&\hspace{0.2cm}T$_{0}$(K)($10^{3}$)\\
      		\hline x=0.0&3.8-14 & 0.038\\
      		x=0.5&3.8-19&0.357\\
      		x=1.0&3.8-12&1.559\\
      		\hline
      	\end{tabular}
      \end{table}
      \vskip 0.5cm
      On the other hand we find that resistivity of all three samples can be fitted to, 3-dimensional Mott variable range 
      hoping (VRH) model \cite{Malinowskia}
      \begin{equation}
      	\rho = \rho_{0}\hspace{0.2cm}exp((T_{0}/T)^{1/4})
      \end{equation}
      where,  T$_{0}$ = $\frac{21.2}{N(E_{F}) \xi^{3}}$, $N(E_F)$  and $\xi$ are density of states at Fermi level and
      localization length respectively. It fits data well but only
      within a limited range of temperature (less than one order of magnitude i.e., from 3 K to 20 K only). 
      T$_{0}$  increases with Nd doping. 
      Assuming that density of states does not vary much between the samples, it shows 
      that $\xi$ decreases 
      with increase in Nd concentration, 
      but since the fit is over such a small range of temperatures,  we do not force this point (strong localization). 
      
       \begin{figure} 
       	\centering
       	\includegraphics[width= 8 cm]{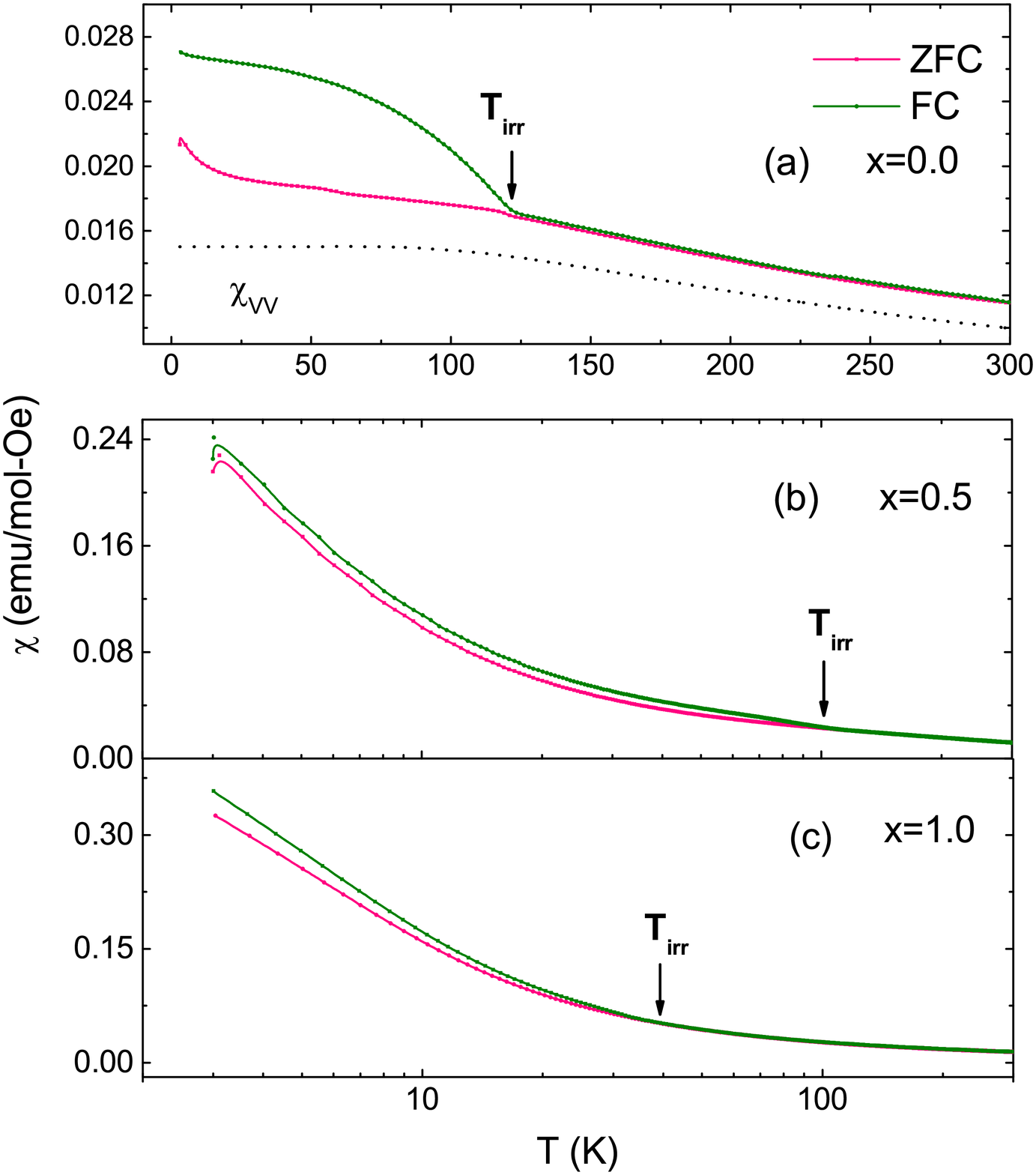}
       	\caption{(a-c)Temperature dependent magnetization measured at 1 KOe in ZFC FC protocol for(Eu$_{1-x}$Nd$_x$)$_{2}$Ir$_{2}$O$_{7}$ where x=0, 0.5, 1.0 and dotted line for x=0 compound represents van vleck suceptibility $ \chi_{VV}$ of Eu$^{3+}$ ion.}
       	\label{MT}
       \end{figure}
       \vskip 0.5cm
        Fig. \ref{MT}(a-c) shows temperature dependent susceptibility  of the samples  in zero field cooled (ZFC)  and field 
        cooled (FC) protocol,  measured under applied magnetic field of 1 kOe within temperature range 3 K - 300 K. M$_{ZFC}$ for parent compound 
        increases with decrease in temperature with  a weak magnetic anomaly close to 120 K. M$_{FC}$ merges with  M$_{ZFC}$ up to 
        temperature 120 K,  below
        which there is a bifurcation between  M$_{ZFC}$ and  M$_{FC}$. The magnetic irreversibility start at T$_{irr}$ $\approx$ 120 K. 
        RXD (Resonant X-ray diffraction) and $\mu$SR (Muon spin rotation and relaxation) measurements 
        have shown that below T$_{irr}$, 
        Ir moments order antiferromagnetically ( all-in/all-out or AIAO),\cite{Eu, Sagayama} though it needs emphasis that
        magnitude of ordered moment is very small. As the interval energy level of Eu$^{3+}$ and Sm$^{3+}$, between the ground state and the successive first excited state is comparable with the thermal energy at the room temperature, they shows temperature dependent Van Vleck suceptibility.\cite{Takikawa} To obtain the contribution 
        of Ir moment in parent compound, 
        we subtracted Van Vleck susceptibility $\chi_{VV}$ due to Eu$^{3+}$ ion from $\chi$.  Temperature dependent 
        $\chi_{VV}$ was calculated 
        by using $\lambda$ = 400 K,\cite{Takikawa} and is shown in fig. \ref{MT}(a) by dotted line.  
        Doped sample also shows bifurcation but lesser in magnitude compared to parent compound. $T_{irr}$  decreases  with 
        Nd concentration.
        T$_{irr}$ for x=0, 0.5 and 1.0 samples are 120 K, 100 K and 38 K respectively and are identical with $T_{MI}$.
        No magnetic anomaly of M$_{ZFC}$ is seen in doped samples down to the lowest temperature.

        %
        %
        
       \begin{figure} 
       	\centering
       	\includegraphics[width= 9 cm]{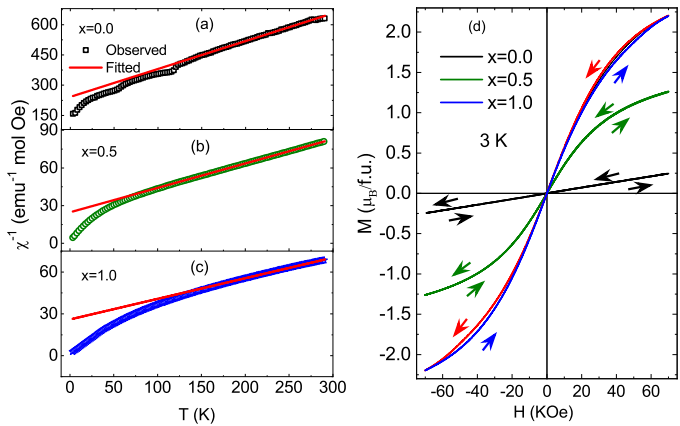}
       	\caption{(a-c)Temperature dependent inverse suceptibility and (d) Isothermal magnetization at 3 K  for (Eu$_{1-x}$Nd$_x$)$_{2}$Ir$_{2}$O$_{7}$ where x=0, 0.5 and 1.0. }
       	\label{Chi-T_M-H}
       \end{figure}
       \begin{table}
       	
       	\centering \caption{$\theta_{p}$, $\mu_{eff}$ for (Eu$_{1-x}$Nd$_x$)$_{2}$Ir$_{2}$O$_{7}$  system.}

       	\begin{tabular}{ccc}
       		\hline Sample&\hspace{0.5cm}$\theta_{p}$ (K)&\hspace{0.5cm}$\mu_{eff}$ ($\mu_{B}$/f.u.)\\
       		\hline x=0.0 &-175 & 2.40\\
       		\hline x=0.5 & -127 & 6.41\\
       		\hline x=1.0 &-175 & 7.35\\
       		
       		\hline
       	\end{tabular}
       \end{table}
        \vskip 0.5cm
        Fig. \ref{Chi-T_M-H}(a-c) shows temperature dependent inverse susceptibility in the entire
        temperature region, fitted in the temperature 
        range 140-290 K with Curie-Weiss law $\chi = \frac{C}{T-\theta{p}}$, $\theta{p}$ is  Curie Weiss  temperature, and Curie constant  
        $C= \frac{N_A\mu_{eff}^{2}}{3K_{B}}$, and  $\mu_{eff}$ is the effective paramagnetic moment. 
        $\theta{p}$, $\mu_{eff}$  for all the samples 
        are shown in table II. At the high temperature region (140-290 K) all the materials show Curie-Weiss
        susceptibility. As the temperature decreases inverse suceptibility deviates from linear dependence with changing its slope below 140 K. At very low temperatures, inverse susceptibility starts  bending down towards low values, presumably because of pure Curie contribution coming from some noninteracting isolated moments. 
        All the sample shows negative $\theta{p}$ at the high temperature region, indicates antiferromagnetic interaction. It is hard to determine the exact moment contribution of Nd$^{3+}$ and Ir$^{4+}$ from this measurement. If we assume the contribution of magnetic moment for Nd$^{3+}$ (3.57 $\mu_B$) from reported Nd$_2$Sn$_2$O$_7$\cite{bertin} where Sn$^{4+}$ has no moment, then the contribution of  Ir local moment in the metallic phase for x=0.5 and 1.0 compounds are 1.42 $\mu_B$ and 0.11 $\mu_B$ respectively. This indicates that Ir moment decreases with Nd doping. Decrease of Ir para-moment in doped materials indicates occurring of strong singlet correlation
        between Ir moments by f-d exchange interaction with Nd moments.

        \vskip 0.5cm
        Isothermal magnetization data (M-H) taken at  3 K,  up to magnetic field 70 kOe of all samples are 
        shown in fig. \ref{Chi-T_M-H}(d). Parent compound shows linear behavior without any magnetic saturation up to 
        magnetic field 70 kOe.  x= 0.5 and 1.0 compounds show non-linear behavior without  any magnetic 
        saturation up to field 70 kOe. Magnetic moment  increases with Nd doping and  moment at  70 kOe,  of x=0.0, 0.5 and 1.0 
        compounds are  0.246 $\mu_{B}$/f.u, 1.263 $\mu_{B}$/f.u and 2.20 $\mu_{B}$/f.u respectively. This low value compared to paramagnetic moments given in table II clearly indicates onset of antiferromagnetic ordering.
        Close view of fig. \ref{Chi-T_M-H}(d) shows  that coercive field (H$_C$) and remanent magetization (M$_r$) for x=0.0 and 0.5 and 1.0 
        compounds are 0.0 Oe, 30 Oe, 20 Oe and 0.0,  4.16$\times10^{-5}$ and 8.29$\times10^{-4}$ $\mu_B$/f.u respectively.

       \vskip 0.5cm        
    
       \begin{figure} 
       	\centering
       	\includegraphics[width= 7 cm]{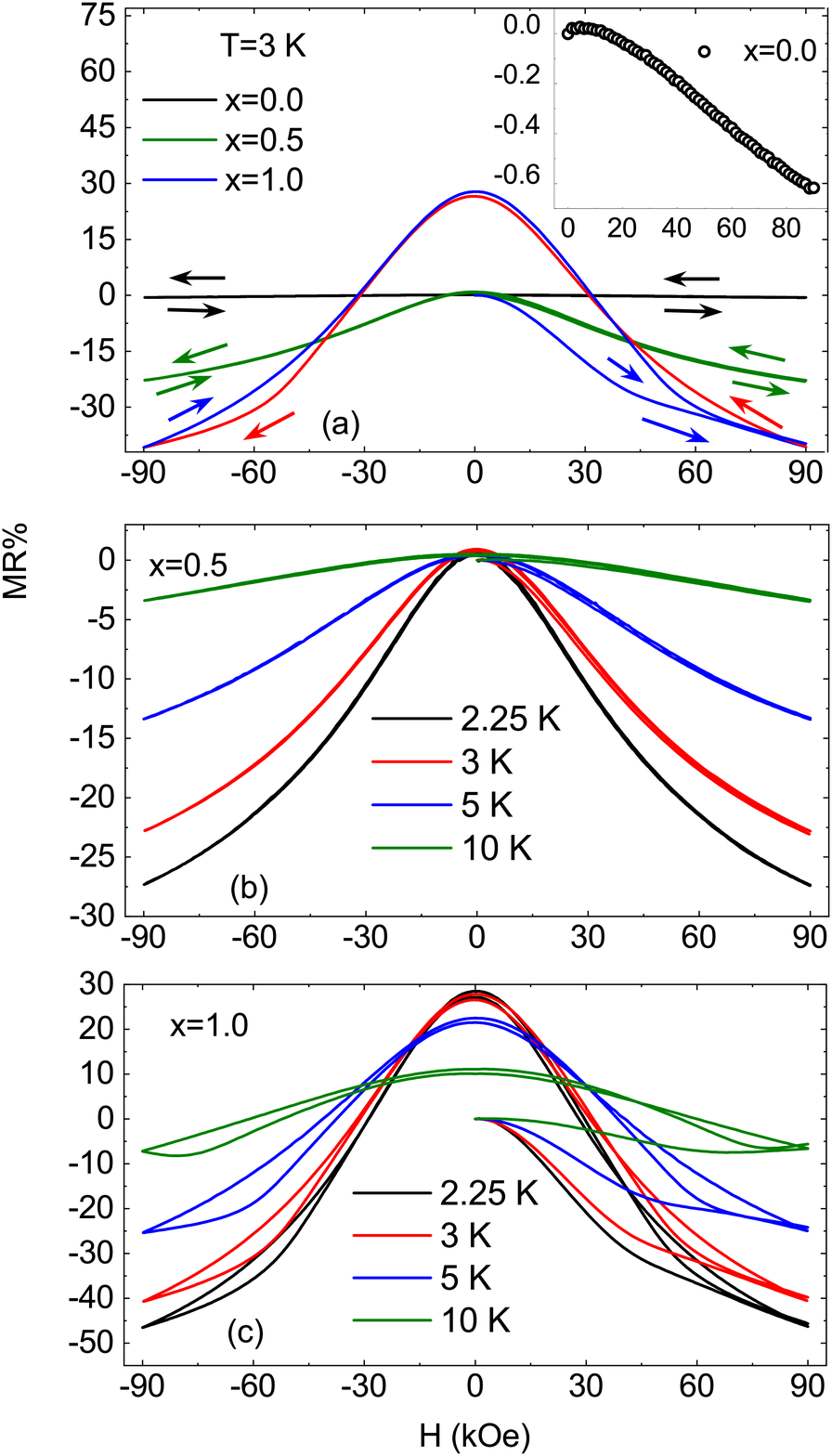}
       	\caption{(a) Magnetic field dependent MR for (Eu$_{1-x}$Nd$_x$)$_{2}$Ir$_{2}$O$_{7}$ where x=0, 0.5 and 1.0, at 3 K with field -90 kOe to +90 kOe and inset shows close view  field dependent MR for x = 0.0 (b, c) Magnetic field dependent MR for x=0.5 and 1.0 compounds at different temperatures.}
       	\label{MR}
       \end{figure}
       
\begin{figure} 
       	\centering
       	\includegraphics[width= 8 cm]{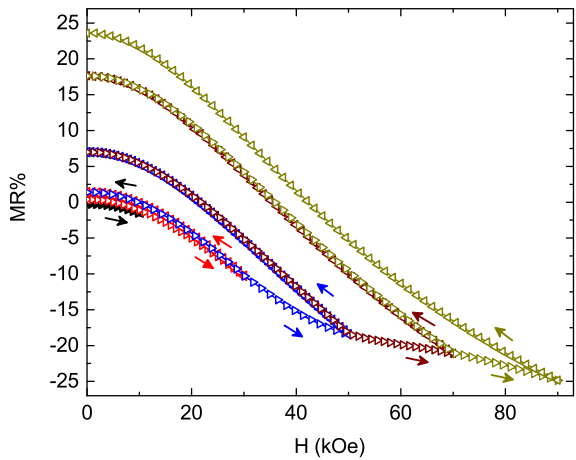}
       	\caption{Different magnetic field cycling MR at 5 K for the compound (Eu$_{1-x}$Nd$_x$)$_{2}$Ir$_{2}$O$_{7}$ where x=1.0.} 
       	\label{MR_relaxation}
       \end{figure}

       	 Magnetoresistance (MR) defined by  [$\rho_{H}$-$\rho_{0}$]/$\rho_{0}$, where $\rho_{0}$ and $\rho_{H}$ are resistivity without and with 
       	applied field are shown in fig. 5(a-c).
       		Fig. \ref{MR}(a) shows magnetic field dependent MR  at 3 K for all the samples. The parent compound shows very small negative
       		MR (0.6 $\%$ at 90 kOe), which varies quadratically at low field (see inset fig. \ref{MR}(a)) and switch to linear field dependency beyond 50 kOe, though Mathsuhira et al.\cite{MATHSUHIRA2} 
       		reports small and positive magnetoresistance at all temperatures. x=0.5 and 1.0 material show  large negative magnetoresistance,
       		varying quadratically at low fields and switching over to linear dependence beyond 50 kOe. Magnitude of
       		negative magnetoresistance keep increasing with decrease in temperature and  at 2.25 K and 90 kOe, reaches 28 $\%$ and 45 $\%$ in
       		x= 0.5 and 1.0 compounds respectively, shown in fig. \ref{MR}(b-c). x=1.0 compound  shows a hysteresis loop between 1st up and down sweep of the field. 
       		This hysteresis loop in 
       		MR for x=1.0 is also accompanied by tiny hysteresis in isothermal magnetization data. It is clear that negative magnetoresistance
       		increases dramatically only with incorporation of 
       		Nd moment and its interaction (f-d exchange) with Ir electrons. 
   Another feature i.e., increase of resistance when the magnetic field is cycled back to zero field, as shown in fig. \ref{MR_relaxation}, is proportional to the magnitude of magnetic field cycle. This has been explained in the discussion section.

       	\begin{figure} 
       		\centering
       		\includegraphics[width= 8.5 cm]{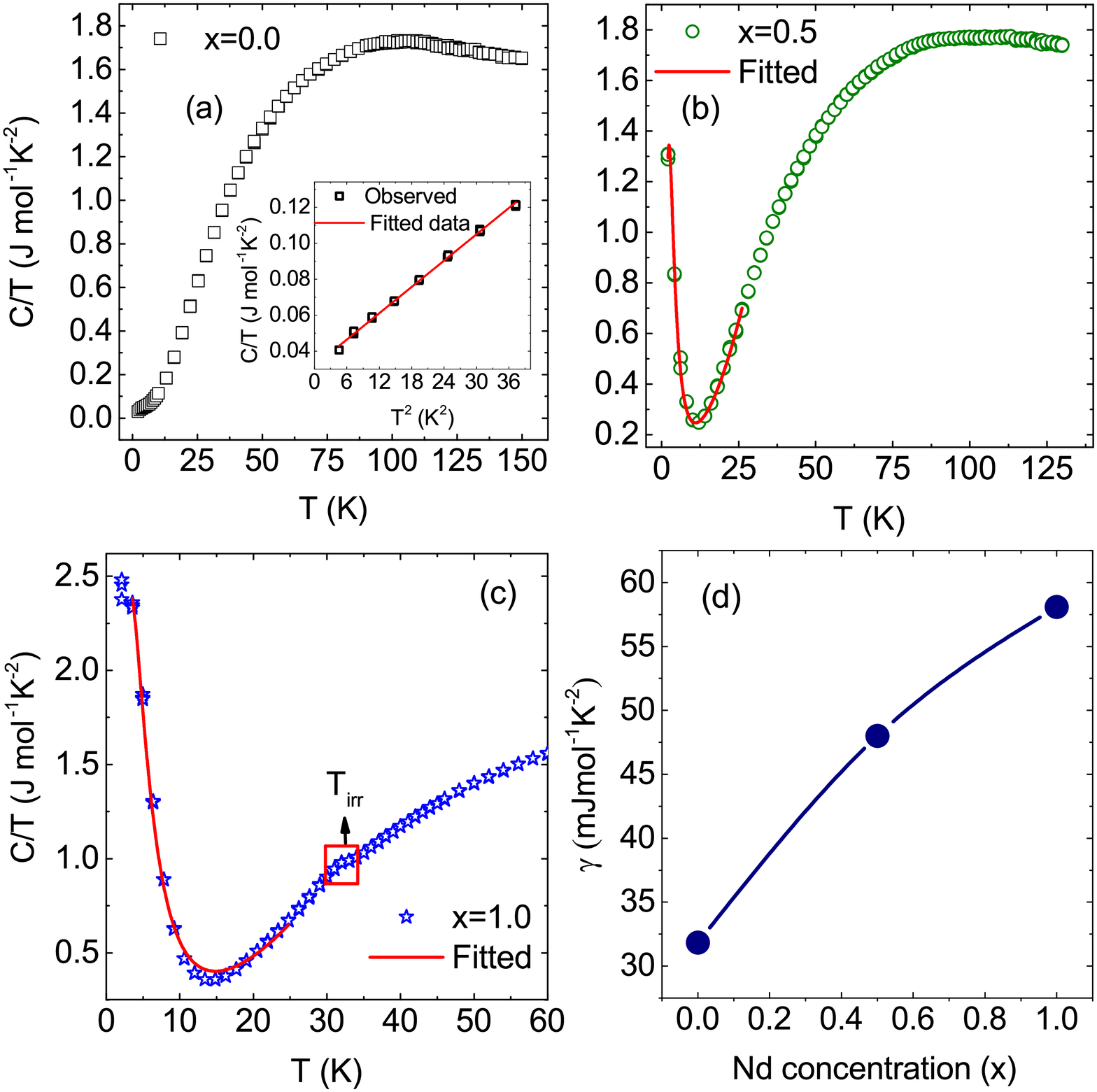}
       		\caption{(a-c) Zero field specific heat (C/T) as a function of temperature between different temperature range for x=0.0, 0.5 and 1.0 compound respectively and solid lines are fitting data. Inset of fig. \ref{HC} (a) shows  C/T vs. T$^2$ data at low temperature for x= 0.0 compound. (d) Variation of $\gamma$ with Nd doping.}
       		\label{HC}
       	\end{figure}
      \vskip 0.5cm  
         Fig. \ref{HC}(a-c) shows temperature variation of  heat capacity in  the  temperature range 2.1-150 K, 2.1-130 K 
         and 2.1-60 K for x=0.0, 0.5 and 1.0 compounds respectively during warming. We observe that there is no anomaly at the T$_{irr}$ for x=0.0 and 0.5 compound though for x=1.0 compound shows  a faint anomaly at 35 K.
         Inset in fig. \ref{HC}(a) shows C/T vs. T$^2$ data for x=0.0 compound.
         We get excellent fit to specific heat within the temperature region 2.1-6 K, by assuming,
         \begin{equation}
         	C = \gamma T + \beta T^3
         \end{equation}
          The fitting parameters  $\gamma$ and $\beta$ are 31.8 mJ mol$^{-1}$ K$^{-2}$ 
         and 0.954 mJ mol$^{-1}$ K$^{-4}$. 
         The  $\gamma$ value is close to  33 mJ mol$^{-1}$ K$^{-2}$ reported
         by Ishikawa et al.\cite{Ishikawa} 
         It is important to point out that a linear in temperature term is
         absolutely necessary to fit the data at lower temperatures ($T < 30$ K). This is surprising because 
         in that temperature region
         the resistivity is non metallic. Weakly localized Fermi liquids can give linear 
         specific heat, because of finite density of states
         near Fermi energy, but the resistivity of parent compound does not follow ${1\over T^{0.5}}$ for a 
         3-dimensional dirty  metal.
         In strongly localized Fermi systems (that shows variable range hopping type of conductivity), specific heat 
         comes purely
         from spin degrees of freedom and is not linear. Moreover in strong localization 
         limit spin susceptibility does not follow Curie
         Weiss form as we observe. Dirac or Weyl fermions with linear energy dispersion gives a $T^3$ specific heat.
         Spin waves (bosons) about antiferromagnetically ordered state also dont give linear specific heat.
         We shall comment on it in discussion section.
         
         In fig. \ref{HC}(b-c) we see  anomalous increase of specific heat at low temperatures,
         in both x=0.5 and 1.0 compounds, that 
         looks like a Schottky type  anomaly.
         We  get good fit by assuming,\cite{Ghosh}

         \begin{equation}
         	C = \gamma T + \beta T^3 + n (\frac{\Delta_{0}}{T})^2\frac{\exp(\Delta_{0}/T)}{(1+\exp (\Delta_{0}/T))^2}
         \end{equation}
         where n  and $\Delta_{0}$ are proportional to the number of two level systems and the energy separation between two levels respectively. 
         

         \begin{figure} 
         	\centering
         	\includegraphics[width= 8 cm]{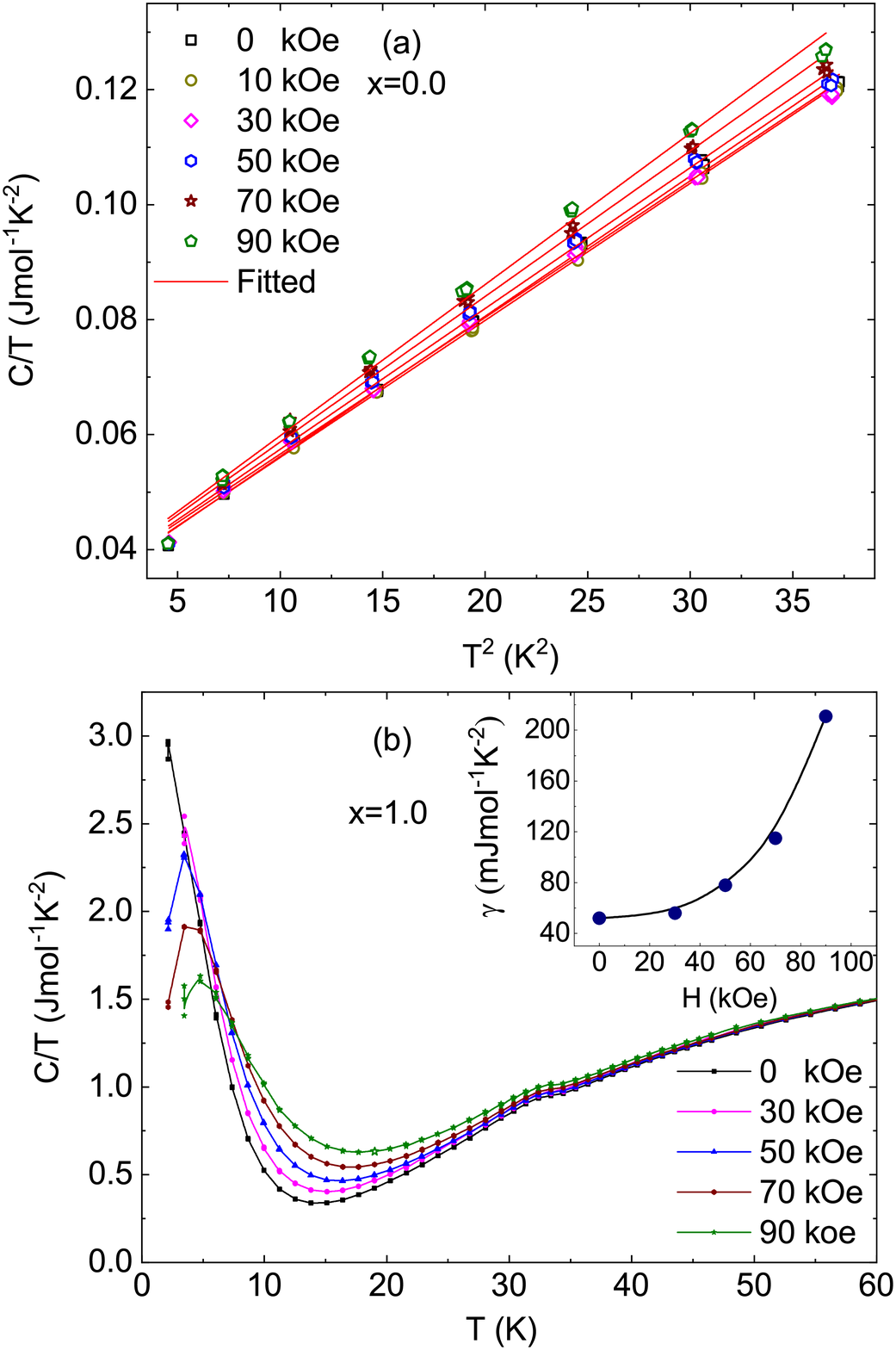}
         	\caption{ (a)  C/T vs. T$^2$ data with different field for x=0.0 compound with fitting. (b) Magnetic field dependent specific heat as a function of temperature for x=1.0 compound. Inset shows field dependent $\gamma$.}
         	\label{HC_FW}
         \end{figure}
         The fitting parameters $\gamma$, $\beta$, $n$ and $\Delta_{0}$ for x=0.5 and 1.0 compounds are 
         48  mJ mol$^{-1}$ K$^{-2}$, 0.945 mJ mol$^{-1}$ K$^{-4}$, 8.2 ,  7.83 K (0.67 meV) and 58.1  
         mJ mol$^{-1}$ K$^{-2}$, 0.907 mJ mol$^{-1}$ K$^{-4}$, 19.82  and 10.35 K (0.89 meV) respectively. We note that the value of $\Delta_{0}$ is very similar to the reported value obtained using inelastic neutron scattering (i.e., 1.3 meV). \cite{Tomiyasu, Watahiki} 
         Coefficient  $\gamma$ increases with Nd doping, but $\beta$ does not vary much.  Both the number of degrees 
         of freedom  $n$ 
         and energy separation between two level $\Delta_{0}$  increases with Nd doping. 
         We suggest that, Schottky anomaly occur because of an effective molecular field H$_{mf}$ via f-d exchange 
         interaction present at the Nd sites once the weak AIAO antiferromagnetic transition of Ir moments sets in as suggested earlier for this system.\cite{Tomiyasu, Watahiki} 
         Nd$^{3+}$ is Krammer ion, and H$_{mf}$ splits the ground state doublet of each Nd$^{3+}$ at low temperature. 
         If the effective moment of Nd$^{3+}$ is $\mu_{Nd}$ then $\Delta_{0}$ = 2$\mu_{Nd}$ H$_{mf}$.
         There will be some additional contribution to H$_{mf}$ from Nd-Nd superexchange interaction in a mean field sense.
         
         %
         
         Fig. \ref{HC_FW}(b) shows that the peak (Schottky) temperature progressively shifts towards higher temperature (i.e., $\Delta_0$ increases with applied magnetic field)
         and  the peak height decreases due to increase in ordering (decrease in entropy) as we increase the external magnetic field from 0.0 to 90 kOe. This is typical
         of Schottky peak coming from localised  two level systems of spin origin (Nd moment).
         
         More interesting is the magnetic field dependence of linear specific heat coefficient $\gamma$ for x=1.0. Fig. \ref{HC_FW}(a) shows that in x=0.0 material
         $\gamma$ does not vary with magnetic field as was observed earlier in spin 1/2 gapless spin liquid.\cite{yamashita1,yamashita2} 
         In x=1.0 material on the other hand $\gamma$ varies superlinearly with magnetic field beyond 30 kOe and shows no sign of
         saturation even at 90 kOe ( see inset in fig. \ref{HC_FW}(b)). 
         Both linear specific heat and its  curious variation with magnetic field  was  not seen before in pyrochlore iridate compounds, 
         and in discussion section we point out the possible origin.
   
     \vskip 0.5cm
     {\large\bf{Discussion}}
     \vskip 0.2cm
     Physics of pyrochlore iridates is dictated by 5d electrons of Ir, which has strong spin-orbit coupling
     and a moderate Hubbard repulsion U (due to large spatial extent of 5d orbitals). Combined effect of octahedral 
     crystal field of oxygen anions and spin-orbit coupling leads to an effective single band (half filled) description 
     in terms of pseudo spin (spin-orbit entangled) $J_{eff}=1/2$ states.

     Resistivity, as we have already discussed has two main problems, (1) inherent disorder due to interchange of
     rare earth and transition metal sites, and any other off-stiochiometry due to oxygen concentration and (2) The observation
     of domain formation and lower resistance in domain boundary region than in the bulk. Therefore transport measurements do not
     reflect precise electronic states in the bulk.

     Both Ir and Nd  ion sublattices are pyrochlore lattice. Superexchange interaction between Ir-Ir and Nd-Nd moments 
     are antiferromagnetic. Moreover 
     due to spin-orbit coupling both Ir and Nd pseudo spin 1/2's have  single ion anisotropy along local [111] axis
     (towards the center of tetrahedras).
     The ground state of such a model is AIAO ordering for both Ir and Nd moments. The f-d exchange interaction 
     has been derived theoretically.\cite{Chen} It is seen that f-d interaction is such that, a local 
     AIAO ordering of Ir moment, gives an effective magnetic field at Nd moments sites encouraging AOAI ordering of
     Nd moments.
     Reason for magnetic irreversibility (field induced moment) is puzzling. It is true that magnetic field can destabilise
     AIAO ordering and lead to 3-in-1-out or 2-in-2-out states, because their energies might be very close to AIAO magnetic state.
     All these states have a net moment per tetrahedra (unlike in AIAO state), but these net moment must alternate going from one
     tetrahedra to the next, and an increase in magnetization induced by magnetic field is not expected.
     We suggest that induced moment comes because of inherent disorder of the kind we mentioned before.
     Suppose 
     a few Ir and Eu ions switch sites (interchange)
     in 
     Eu$_2$Ir$_2$O$_7$.
     Then If an  Eu$^{3+}$  ion ( zero net pseudospin) sits at an Ir site,
     then Ir moments in all tetrahedras connected to this site will deviate from AIAO ordering, i.e. those tetrahedras
     will pick up a net magnetic moment. Secondly Ir electron from a near neighbour site can now hop into 
     empty 5d orbital of Eu or Nd ions (delocalisation of Ir electrons through Eu/Nd on Ir sublattice  sites), creating an effective 
     double exchange type local ferro correlation along with a net moment.

     As the material  cools down
     from high 
     temperature, slowly Ir-Ir superexchange and magnetic anisotropy takes over and Ir moments 
     starts ordering
     (AIAO) below T=120 K, 
     though the observed  magnitude of AIAO ordered Ir moments are very small.\cite{Tomiyasu} 
     The AIAO ordering of Ir moments gives an effective magnetic field at Nd moment sites. This is visible from the 
     observed Schottky peak in the specific heat data for x=0.5 and 1.0 compound. However, the inverse suceptibility tending towards zero value in fig. \ref{Chi-T_M-H}(a-c) indicates a paramagnetic background at low temperature.

     It has been argued\cite{fujita}  that real space spin chirality, defined as $\chi_{ijk} = S_i \cdot S_j \times S_k$ is responsible
     for negative magnetoresistance. Spin chirality\cite{wen} can be written in terms of underlying electron operators (fermions)
     as, $ { i\over 2} \chi_{ijk} = P_{ijk} -  P_{ikj}$, where $P_{ijk}= \chi_{ij} \chi_{jk} \chi_{ki}$ and 
     $\chi_{ij}= C_{i\sigma}^{\dagger} C_{j \sigma} $, where $C$'s are electron operators.
     A non zero value of $\chi_{ijk}$  leads to a Berry  phase picked up by electrons as it  circulates around a triangular
     plaquette, i.e as if spin chirality creates a fictitious magnetic field perpendicular to the plaquette.
     This chirality has to be contrasted with chirality of Weyl fermions near Weyl points, that are created by a singular Berry phase
     in momentum space.\cite{nagaosa} This kind of real space spin chirality can give  Hall effect without magnetic field
     as observed in MnSi\cite{taguchi} and negative magnetoresistance.\cite{neu} When spin chirality of
     all triangular plaquettes are staggered or along random direction, the magnetoresistance varies quadratically with magnetic field.
     At higher magnetic field when the chirality of most triangular plaquettes are preferably aligned along some
     direction then asymmetric scattering of electrons by spin chirality can occur and one gets odd-parity magnetoresistance, i.e magnetoresistance varying linearly with magnetic field. Therefore the magnetic ordering of Ir electrons can be thought of
     as a weak antiferromagnetic (AIAO) long range ordering (Reported earlier by RXD and $\mu$SR measurement\cite{Eu,Sagayama}) superimposed on a chiral spin liquid state. \cite{Machida,Nagata,Zorko,Kenney}


     Huge increase in magnetoresistance with increase in Nd, points out importance of Nd moments.
     We suggest that, apart from the scattering of carriers in the domain walls by chirality fluctuations,
     there is an additional Kondo scattering of carriers by Nd electron moments. Kondo scattering in heavy fermion materials
     give quadratic field dependent at low field and linear at high field negative magnetoresistance.\cite{hewson} For Nd doping the resistance value increases after a field cycle is due to reduction of domain wall i.e., in between either AIAO-AOAI or vice versa ordered regions.\cite{eric}
     This was noticed before in Pr$_2$Ir$_2$O$_7$ \cite{balicas} and in Nd$_2$Ir$_2$O$_7$. \cite{MATSUHIRA1}

     Now we come to our observation of linear specific heat in all three samples at low temperature insulating
     phase.  This is most unusual. Linear specific heat can occur in Fermi system in weak localization regime,
     because of non zero density of states near the Fermi energy, but like we discussed earlier the temperature dependent resistivity
     rules out weak localisation. Weyl fermions show non-metallic resistivity (negative temperature coefficient of
     resistivity) but because of their linear dispersion, Weyl fermions would rather give a $T^3$ specific heat.
     Antiferromagnets in frustrated lattices often give spin-liquid ground states
     supporting  exotic  excitations called spinons obeying fractional 
     quantum statistics.\cite{balent} 
     Theoretical\cite{ran} analysis of Kagome lattice antiferromagnet suggests that,  a  linear specific heat in insulating antiferromagnet
     is possible if spinons have finite area Fermi surface. Spin liquids are  like spin superfluids (pairing between
     spinons of opposite spins) with or without gap for spinon excitations.\cite{bas} Free spinons give linear specific heat.
     In x=0.0  material $\gamma$ does not change with the magnetic field and this is consistent with a spin liquid state
     having gapless spinon excitations\cite{yamashita1,yamashita2} coexisting with long range antiferromagnetic ordering\cite{Sagayama} as is observed in other magnetically ordered systems.\cite{Nagata,Zorko,Kenney} The magnitude of $\gamma$ in x=1.0 compound is also much larger (this is due to contribution of spinons from both the Ir and Nd sublattice ) than
          in x=0.0 compound in zero field. 
          It is also possible that Nd spin system develops spin liquid kind of singlet formation due to f-d exchange interaction. The f-d exchange interaction gets modified beyond a certain applied magnetic field which induces breaking of Nd and Ir spin singlets.
     Hence, in x=1.0 , $\gamma$ varies superlinearly beyond certain magnetic field and do not show any sign of saturation even 
     at  90 kOe (see inset of fig. \ref{HC_FW}(b)).
     This suggests that spin liquid type singlet correlation is in both Ir and Nd spin sub-systems without affecting the long range antiferromagnetic ordering. Here we would like to point out categorically that there is coexistence of Neel order (AIAO)\cite{Guo,Nd} and singlet spin liquid order (i.e., long range resonating valence bond ordering).\cite{Nagata,Zorko,Kenney}
     It needs more experimental investigation of two pyrochlore antiferromagnets coupled by f-d exchange. 
     
     \vskip 0.5cm
     {\large\bf{Conclusion}}
     \vskip 0.2cm
      From different magnetic field cycling MR it is clear that these materials are fragmented into domains 
      with larger resistivity (larger charge gap) separated by domain walls with lesser resistivity (lesser charge gap).
      From our  analysis of magnetic susceptibility, low temperature
      linear specific heat and negative magnetoresistance with low field quadratic and high field linear dependence we conclude  that,  the magnetic state of Ir electrons in the domains (larger volume fraction),
      is very weak antiferromagnetic
      AIAO ordering superimposed on  a predominatly possible chiral spin liquid state, i.e coexistence of AIAO ordering and chiral spin liquid.
      Observation of linear specific heat in the insulating low temperature phase rules out the possibility of Weyl semimetal
      phase in these materials. Linear specific heat comes from spinons, which are excitations above
      the spin liquid state. We also conclude that f-d exchange interaction encourages singlet correlation (spinon pairing).
      Transport properties like resistivity and magnetoresistance comes from the domain wall regions, where  small AIAO
      ordering of Ir moments is destroyed, leading to smaller  charge gap than in the domains. 
      In x=0.0 material,  this chiral spin
      liquid gives very small negative magnetoresistance.  In x=0.5 and 1.0 materials  magnetoresistance is hugely amplified
      by Kondo scattering of  rare earth moments with the chiral spin liquid.  
      Chiral spin liquid along with
      Kondo scattering could be  exciting areas that might lead to new physics. 
      
     \vskip 0.5cm
     {\large\bf{Acknowledgement}}
     \vskip 0.2cm
     One of the author SM would like to thank Anish Karmahapatra, ECMP Division, SINP for XRD measurements. This work is partially supported by SERB, DST, GOI under TARE project (File No.:TAR/2018/000546).
     
    \end{document}